\documentstyle[graphicx]{mn}

\newif\ifAMStwofonts

\def\lsim{\raise0.3ex\hbox{$<$}\kern-0.75em{\lower0.65ex\hbox{$\sim$}}}
\def\gsim{\raise0.3ex\hbox{$>$}\kern-0.75em{\lower0.65ex\hbox{$\sim$}}}
\def\msun{M_\odot}
\def\lbrack2{[\![}
\def\rbrack2{]\!]}

\def\1D{one spatial dimension}
\def\2D{two spatial dimensions}
\def\3D{three spatial dimensions}

\def\dt{{{\small \Delta}t}}

\def\xe{{x_{\rm e}}}

\def\km{{\rm\,km}}

\def\kpc{{\rm\,kpc}}
\def\mpc{{\rm\,Mpc}}
\def\msun{{\rm\,M_\odot}}

\def\cm{{\rm\,cm}}
\def\s{{\rm\,s}}
\def\erg{{\rm\,erg}}
\def\ster{{\rm\,sr}}
\def\hz{{\rm\,Hz}}
\def\hp{h_{\rm P}}

\def\ev{{\rm\,eV}}
\def\yrs{{\rm\,yrs}}

\newcommand\comment[1]{\null}
\newcommand\hii{\hbox{H\,{\sc ii}~}}

\ifoldfss
  \ifCUPmtlplainloaded \else
    \NewTextAlphabet{textbfit} {cmbxti10} {}
    \NewTextAlphabet{textbfss} {cmssbx10} {}
    \NewMathAlphabet{mathbfit} {cmbxti10} {} 
    \NewMathAlphabet{mathbfss} {cmssbx10} {} 
  \fi
  \ifAMStwofonts
    \ifCUPmtlplainloaded \else
      \NewSymbolFont{upmath} {eurm10}
      \NewSymbolFont{AMSa} {msam10}
      \NewMathSymbol{\upi}     {0}{upmath}{19}
      \NewMathSymbol{\umu}     {0}{upmath}{16}
      \NewMathSymbol{\upartial}{0}{upmath}{40}
      \NewMathSymbol{\leqslant}{3}{AMSa}{36}
      \NewMathSymbol{\geqslant}{3}{AMSa}{3E}

      \let\leq=\leqslant \let\le=\leqslant
       
    \fi
  \fi
\fi 

\ifnfssone
  \newmathalphabet{\mathit}
  \addtoversion{normal}{\mathit}{cmr}{m}{it}
  \addtoversion{bold}{\mathit}{cmr}{bx}{it}
  \newmathalphabet{\mathbfit} 
  \addtoversion{normal}{\mathbfit}{cmr}{bx}{it}
  \addtoversion{bold}{\mathbfit}{cmr}{bx}{it}
  \newmathalphabet{\mathbfss} 
  \addtoversion{normal}{\mathbfss}{cmss}{bx}{n}
  \addtoversion{bold}{\mathbfss}{cmss}{bx}{n}
  \ifAMStwofonts
    \ifCUPmtlplainloaded \else
      \UseAMStwoboldmath
      \makeatletter
      \new@mathgroup\upmath@group
      \define@mathgroup\mv@normal\upmath@group{eur}{m}{n}
      \define@mathgroup\mv@bold\upmath@group{eur}{b}{n}
      \edef\UPM{\hexnumber\upmath@group}
      \new@mathgroup\amsa@group
      \define@mathgroup\mv@normal\amsa@group{msa}{m}{n}
      \define@mathgroup\mv@bold\amsa@group{msa}{m}{n}
      \edef\AMSa{\hexnumber\amsa@group}
      \makeatother
      \mathchardef\upi="0\UPM19
      \mathchardef\umu="0\UPM16
      \mathchardef\upartial="0\UPM40
      \mathchardef\leqslant="3\AMSa36
      \mathchardef\geqslant="3\AMSa3E

      \let\leq=\leqslant \let\le=\leqslant

    \fi
  \fi
\fi 

\ifnfsstwo
  \DeclareMathAlphabet{\mathbfit}{OT1}{cmr}{bx}{it}
  \SetMathAlphabet\mathbfit{bold}{OT1}{cmr}{bx}{it}
  \DeclareMathAlphabet{\mathbfss}{OT1}{cmss}{bx}{n}
  \SetMathAlphabet\mathbfss{bold}{OT1}{cmss}{bx}{n}
  \ifAMStwofonts
    \ifCUPmtlplainloaded \else
      \DeclareSymbolFont{UPM}{U}{eur}{m}{n}
      \SetSymbolFont{UPM}{bold}{U}{eur}{b}{n}
      \DeclareSymbolFont{AMSa}{U}{msa}{m}{n}
      \DeclareMathSymbol{\upi}{0}{UPM}{"19}
      \DeclareMathSymbol{\umu}{0}{UPM}{"16}
      \DeclareMathSymbol{\upartial}{0}{UPM}{"40}
      \DeclareMathSymbol{\leqslant}{3}{AMSa}{"36}
      \DeclareMathSymbol{\geqslant}{3}{AMSa}{"3E}

      \let\leq=\leqslant \let\le=\leqslant

    \fi
  \fi
\fi 

\ifCUPmtlplainloaded \else
  \ifAMStwofonts \else 
    \def\upi{\pi}
    \def\umu{\mu}
    \def\upartial{\partial}
  \fi
\fi

\title{3D numerical cosmological radiative transfer in an inhomogeneous medium}
\author[A.~O.~Razoumov and D.~Scott]
	{Alexei O.~Razoumov\thanks{e-mail:  razoumov@astro.ubc.ca}
	and Douglas Scott\thanks{e-mail: dscott@astro.ubc.ca}\\
	Department of Physics \& Astronomy,
        University of British Columbia, Vancouver B.C.~V6R 1Z4,~~Canada}

\date{original form Oct. 1998}

\pagerange{\pageref{firstpage}--\pageref{lastpage}}
\pubyear{1998}

\begin{document}

\maketitle

\label{firstpage}

\begin{abstract}
A numerical scheme is proposed for the solution of the three-dimensional
radiative transfer equation with variable optical depth. We show that {\it
time-dependent} ray tracing is an attractive choice for simulations of
astrophysical ionization fronts, particularly when one is interested in
covering a wide range of optical depths within a 3D clumpy medium. Our
approach combines the explicit advection of radiation variables with the
implicit solution of {\it local} rate equations given the radiation field at
each point. Our scheme is well suited to the solution of problems for which
line transfer is not important, and could, in principle, be extended to
those situations also. This scheme allows us to calculate the propagation
of supersonic ionization fronts into an inhomogeneous medium. The approach
can be easily implemented on a single workstation and also should be fully
parallelizable.
\end{abstract}


\section{Introduction}
\label{sec:intro}

Understanding the effect of the radiation field on the thermal state
of interstellar and intergalactic gas is important for many areas of
astrophysics, and in particular for star and galaxy formation. Of
special interest for cosmological structure formation are the epoch of
reionization of the Universe (Gnedin \& Ostriker~1997,
Miralda-Escud\'e et al.~1998), the effects of self-shielding on the
formation of disk and dwarf galaxies (Navarro \& Steinmetz~1997,
Kepner et al.~1997) and the absorption properties of Ly$\,\alpha$
clouds (Katz et al.~1996, Meiksin 1994).  While our knowledge of the
physics of large-scale structure and galaxy formation has benefited
significantly from numerical N-body and gas-dynamical models (see,
e.g., Zhang et al.~1998, and references therein), there is very little
that has been done to include radiative transfer (RT) into these
simulations. Challenges seem to abound, not least of all, the fact
that the intensity of radiation in general is a function of seven
independent variables (three spatial coordinates, two angles,
frequency and time). While for many applications it has been possible
to reduce the dimensionality (e.g.~to build realistic stellar
atmosphere models), the clumpy state of the interstellar or
intergalactic medium does not provide any spatial
symmetries. Moreover, coupled equations of radiation hydrodynamics
(RHD) have very complicated structure, and are often of mixed
advection-diffusion type which makes it very difficult to solve them
numerically. Besides that, the radiation field in optically thin
regions usually evolves at the speed of light, yielding an enormous
gap of many orders of magnitude between the characteristic time-scales
for a system.

One way to avoid the latter problem is to solve all equations on the
fluid-flow time-scale. While there are arguments which seem to
preserve causality in such an approach \cite{mihalas84}, even then the
numerical solution is an incredibly difficult challenge
\cite{stone92b}. On the other hand, several astrophysical problems
allow one to follow the system of interest on a radiation propagation
time-scale, without imposing a prohibitively large number of time
steps. In the context of cosmological RT, we would like to resolve the
characteristic distance between the sources of reionization. Cold
dark matter (CDM) cosmologies predict the collapse of the first baryonic
objects as early as $z=30-50$, with the typical Jeans mass of order
$10^5 \msun$ (see Haiman \& Loeb~1997, and references therein). The
corresponding comoving scale of fragmenting clouds is $\sim 7\kpc$.
If stellar sources reside in primordial globular clusters of this
mass, then for $\Omega_{\rm b}=0.05$ the average separation between these
objects is $\Delta x \sim 20\kpc$ (comoving). For explicit schemes the
Courant condition imposes a time-step

\begin{equation}
\label{eq:cfl}
\dt \le t_{\rm R}={\Delta x\over c(1+z)} \sim {10^5 \over 1+z} h^{-1} \yrs.
\end{equation}

\noindent
Thus evolution from $z\,{=}\,20$ to $z\,{=}\,3$ takes $\sim7.5\times
10^8 h^{-1} \yrs$ (assuming density parameter $\Omega_0=1$, and
defining the Hubble constant to be $H_0=100\,h{\rm km}\,{\rm
s}^{-1}{\rm Mpc}^{-1}$). Substituting $z\,{=}\,3$ and $z\,{=}\,20$
into eq.~(\ref{eq:cfl}) yields the total number of time-steps in the
range $20{,}000-150{,}000$ over the entire course of evolution. In
other words, to resolve reionization by $10^5\msun$ stellar clusters,
we need to compute $\sim$ {\it few tens of thousands} of time-steps on
average, and for $10^8\msun$ clusters the number of steps required
will be ten times smaller. A $100^3$ grid with the required resolution
will result in computational boxes of several $\mpc$ on a side. A full
cosmological radiative transfer simulation with boxes at least this
big and for all the required timesteps has not been feasible in the
past.

This is far from the only challenge we face. Since any two points can
affect each other via the radiation field, even for a monochromatic
problem, we must describe the propagation of the radiation field
anisotropies in the full five-dimensional space. Standard steady-state
RT solvers, which have been widely used in stellar atmosphere models,
are not efficient in this case. Non-local thermodynamic equilibrium
(NLTE) steady-state radiative transport relies on obtaining the
numerical solution via an iterative process for the whole
computational region at once, and is usually effective only for very
simplified geometries. Any refinement of the discretization grid
and/or increase in the number of atomic rate equations to compute NLTE
effects will necessarily result in an exponential increase in the
number of iterations required to achieve the same accuracy. On the
other hand, the 3D solution of the steady-state transfer equation in
the absence of any spatial symmetries can often be obtained with Monte
Carlo methods \cite{park98}. However, these methods demonstrate very
slow convergence at higher resolutions and are hardly applicable if
one is interested in following a time-dependent system.

The change in the degree of ionization in a low-density environment
occurs on a radiation propagation time-scale $t_{\rm R}$. To track
ionization fronts (I-fronts) in this regime, it is best to apply a
high-resolution shock-capturing scheme similar to those originally
developed in fluid dynamics. One possible approach is the direct
numerical solution of the monochromatic photon Boltzmann equation in
the 5-dimensional phase space (Razoumov, in preparation). To allow
for a trade-off between calculational speed (plus memory usage) and
accuracy, a more conventional approach is to truncate the system of
angle-averaged radiation moment equations at a fixed moment and to use
some closure scheme to reconstruct the angle dependence of the
intensity at each point in 3D space. The method of variable Eddington
factors first introduced by Auer \& Mihalas (1970) has been shown to
produce very accurate closure for time-dependent problems in both 2D
\cite{stone92b} and 3D \cite{umemura98}. However, to the best of our
knowledge, all schemes employed so far for calculating the
time-dependent variable Eddington factor were based on a steady-state
reconstruction of the radiation field through all of the computational
region at once, given the thermal state of material and level
populations at each point. Since advection (or spatial transport)
of moments is still followed on the time-scale of typical changes in
the ionizational balance of the system, this approximation certainly
provides physically valid results, assuming that the reconstruction is
being performed often enough. However, the steady-steady closure
relies on the iterative solution of a large system of non-linear
equations, which becomes an exceedingly difficult problem, from the
computational point of view, as one moves to higher spatial and
angular resolution and to the inclusion of more complicated
microphysics.

The goal of the present paper is to demonstrate that in the
cosmological context it is possible and practical to solve the whole
RT problem on a radiation propagation time-scale $t_{\rm R}$ -- as
opposed to the fluid-flow time-scale -- and we present a simple
technique which gives an accurate solution for the angle-dependent
intensity in three spatial dimensions. The scheme can track
discontinuities accurately in 3D and is stable up to the Courant
number of unity. Since all advection of radiation variables is being
done at $t_{\rm R}$, the scheme is well tailored to the numerical
study of the propagation of I-fronts into a non-homogeneous medium
with any optical depth, and gives very accurate results for
scattering. Rather than solve the proper chemistry equations
applicable to cosmological structures, we adopt a simple toy model
described below. Similarly, in the current paper we do not make any
attempt to model the thermal properties of the gas, concentrating just
on efficient multidimensional advection techniques.

This paper is organized as follows. In Section~\ref{sec:problem} we
briefly review the state of numerical RT in the study of reionization.
We then concentrate on methods for 5D numerical advection. In
Section~\ref{sec:technique} we describe our numerical algorithm and we
present the results of numerical tests in Section~\ref{sec:tests}.
Finally, in Section~\ref{sec:conclude} we discuss the next steps
towards a realistic 3D RT simulation.

\section{Formulation of the problem}
\label{sec:problem}

It is believed that light from the first baryonic objects \comment {the
first generation of stars and quasars} at $z\ga6$ led to a phase-like
transition in the ionizational state of the Universe. This process of
reionization significantly affected the subsequent evolution of structure
formation (Couchman \& Rees~1986). In detail reionization did not happen at
a single epoch, with details of `pre-heating', percolation, helium
ionization and other physical processes having been studied in great detail
over the last decade (some recent contributions include Madau et al.~1997,
Haiman \& Loeb~1997, Gnedin \& Ostriker~1997, Shapiro et al.~1998, Tajiri \&
Umemura~1998).

It now seems clear that the full solution of the problem requires a
detailed treatment of the effects of RT. To complicate matters, by
the time of the first star formation, the small-scale density
inhomogeneities had entered the non-linear regime \cite{gnedin97},
and the medium was filled with clumpy structures. The success of
cosmological N-body and hydrodynamical models in quantifying the
growth of these objects (e.g., Zhang et al.~1998) suggests that the
next step will be to include the effects of global energy exchange by
radiation.
Indeed, there is a need for time-dependent 3D RT models as numerical
tools for understanding the effect of inhomogeneities in the dynamical
evolution of the interstellar/intergalactic medium. For instance, the
ability of gas to cool down and form structures depends crucially on
the ionizational state of a whole array of different chemical
elements, which in turn directly depends on the local energy density
of the radiation field.

The hydrogen component of the Universe is most likely ionized by
photons just above the Lyman limit, because (1) the cross-section of
photoionization drops as $\nu^{-3}$ at higher frequencies, and (2) the
medium will be dominated by softer photons, even in the case of quasar
reionization (when ionizing photons come mostly from diffuse \hii
regions). Therefore, we argue that either monochromatic or
frequency-averaged transfer will be a fairly good approximation in our
models.

Recently, the problem of simulating 3D inhomogeneous reionization with
realistic radiative transfer has attracted considerable interest in the
scientific community. Umemura et al. (1998) calculated reionization
from $z\,{=}\,9$ to $z\,{=}\,4$, solving the 3D steady-state RT
equation along with the time-dependent ionization rate equations for
hydrogen and helium. The radiation field was integrated along spatial
dimensions using the method of short characteristics
\cite{stone92b}. The steady-state solution implies the assumption that the
radiation field adjusts instantaneously to any changes in the ionization
profile. One draw-back of this approach, however, is in low-density
voids where there are probably enough Lyman photons to ionize every
hydrogen atom, so that the velocity of I-fronts is simply equal to the speed
of light. Then the rate equations still have to be solved on the radiation
propagation timescale. Besides, implicit techniques in the presence of
inhomogeneities will become exponentially complicated, if we want to solve
time-dependent rate equations for multiple chemical species.

Norman et al. (1998) and Abel et al. (1998) present a scheme for
solving the cosmological radiative transfer problem by decomposing the
total radiation field into two parts: highly anisotropic direct
ionizing radiation from point sources such as quasars and stellar
clusters, and the diffuse component from recombinations in the
photoionized gas. In their method the direct ionizing radiation is
being attenuated along a small number of rays, each of which is forced
to pass through one of the few point sources within the simulation
volume. The diffuse part of the radiation field is found with a
separate technique which can benefit from the nearly isotropic form of
this component, for instance, through the use of the diffusion
approximation. Both solutions are obtained neglecting the time
dependent term in the radiative transfer equation, with the default
time-step dictated by the speed of the atomic processes.


If reionization by quasars alone is ruled out (Madau~1998, however see
Haiman \& Loeb~1998), then I-fronts will be caused by Lyman photons
from low-luminosity stellar sources at high redshifts. In this case the
pressure gradient across the ionization zone is more likely to become
important before the front is slowed down by the finite recombination time.
In the present paper we ignore hydrodynamical effects, concentrating on an
efficient method to track supersonic I-fronts. Our approach is to solve the
time-dependent RT coupled with an implicit local solver for the rate equations.
This method gives the correct speed of front propagation and it also quickly
converges to a steady-state solution for equilibrium systems. However, we
should note that until a detailed comparison is made between explicit
advection (at the speed of light) and the implicit reconstruction (through
an elliptic solver), it is difficult to judge which approach works best in
simulating inhomogeneous reionization in detail.

Although radiation propagates with the speed of light and the intensity of
radiation depends on five spatial variables, plus frequency and time, the RT
equation is inherently simpler than the equations of compressible
hydrodynamics, since its advection part is strictly linear. Non-linearities
are usually introduced when we are trying to reduce the dimensionality of
the problem. Much of the difficulty thus comes from inability to get
decent numerical resolution in the 5D (or 6D with frequency) space with
present-day computers.

In the current work we have attempted to develop an efficient method to
describe the anisotropies in the monochromatic radiation field propagating
through an inhomogeneous medium, which we now describe.

\section[]{The numerical technique}
\label{sec:technique}


The classical RT equation (without cosmological terms) reads

\begin{equation}
\label{eq:rte}
{1\over c}{\partial I_{\nu}\over \partial t}+\bmath{n}\cdot \bmath{\nabla}
 I_{\nu}= \epsilon_{\nu}-\kappa_{\nu}I_{\nu},
\end{equation}

\noindent
where $I_{\nu}$ is the intensity of radiation in direction $\bmath{n}$ and
$\epsilon_{\nu}$ and $\kappa_{\nu}$ are the local emissivity and opacity.
This equation is valid for those problems in which the light-crossing
time across the computational volume $L/c$ is much smaller than the Hubble
time, and we can neglect the redshift effects within the simulation volume.
With proper boundary conditions one can easily account for the Doppler shift
in the background radiation.

The basic idea of our technique is to solve eq.~(\ref{eq:rte})
directly for the angle-dependent intensity $I({\bf r},{\bf n},\nu,t)$
at each point. The total radiation field at each point is divided into
the direct (from ionizing sources) and diffuse (due to
recombinations in the gas) components:

\begin{equation}
E_{\nu,i,j,k}=E^{\rm src}_{\nu,i,j,k}+E^{\rm diff}_{\nu,i,j,k},
\end{equation}

\noindent
where $i,j,k$ are the three indices in a rectangular grid. The
energy density $E^{\rm src}_{\nu,i,j,k}$ due to direct photons coming
from point sources of ionization can be easily calculated on the 3D
grid via summation over all sources (assuming that there are not too
many of them within the volume):

\begin{displaymath}
\label{eq:dirph}
E^{\rm src}_{\nu,i,j,k}=\sum_{\rm s}
{h\nu \dot{N}_{\rm ph}\over 4\pi c r_{i,j,k,{\rm s}}^2}
e^{-\tau_{\nu,i,j,k,{\rm s}}}\times
\left\{\begin{array}{ll}
1 & \textrm{if $r_{i,j,k,{\rm s}}\le ct_{\rm s}$}\\
0 & \textrm{if $r_{i,j,k,{\rm s}}> ct_{\rm s}$},
\end{array} \right.
\end{displaymath}

\[
{\rm where}~~~
\tau_{\nu,i,j,k,{\rm s}}=\int_0^{r_{i,j,k,{\rm s}}}d\tau_{\nu}
\]

\noindent
is the optical depth, $r_{i,j,k,{\rm s}}$ is the physical distance
between the current point and the source and $t_{\rm s}$ is the age of
the source. Note that the rate of emission of photons $\dot{N}_{\rm
ph}$ can be modified to allow for variability of sources on short
timescales. For the diffuse component we use an upwind monotonic
scheme to propagate 1D wavefronts $I({\bf r},{\bf n},\nu,t)$ along a
large number of rays in 3D at the speed of light. Following Stone \&
Mihalas (1992), we apply an operator split explicit-implicit scheme,
in which advection of radiation variables is treated explicitly and
the atomic and molecular rate equations are solved implicitly and
separately at each point. Unlike Abel et al.~(1998), we use rays to
track the {\it diffuse} component of the radiation field and the
direct ionizing background radiation (streaming into the computational
volume). Since we need to draw rays
essentially through every grid point in the 3D volume, at first glance
this approach appears to have very large memory requirements. However,
efficient placing of the rays can significantly reduce the
computational effort.

\subsection{A uniform and isotropic grid of rays}

At each time-step we are interested in getting a solution for the mean
radiation energy density and material properties on a 3D $N^3$
rectangular grid. Instead of shooting rays though each grid node in
3D, we choose to cover the whole computational volume with a separate
grid of rays which is uniform both in space and in angular directions.
Assuming that the computational volume corresponds to the range $0\le
x,y,z\le 1$, we first construct a 2D rectangular base grid containing
$N^2$ nodes with coordinates with the origin at the centre of the cube

\[
x_{ij}=\sqrt{3}\left({i-1/2\over N}-{1\over 2}\right),
\]
\[
y_{ij}=\sqrt{3}\left({j-1/2\over N}-{1\over 2}\right),
\]
\begin{equation}
\label{eq:base_grid}
z_{ij}={1\over 2},~~~~~{\rm with}~~i, j=1,\ldots,N,
\end{equation}

\noindent
where the $\sqrt{3}$ factor ensures that the entire volume is covered
with rays, and we shoot rays normal to the base grid through all of
its grid points. Note that the separation between 2D nodes is allowed
to be larger than $1/N$.
To cover the whole volume with rays, we then rotate the base grid by
an angle $\theta_{l}-\pi/2$ around the $y$-axis and by $\phi_{lm}$
around the $z$-axis, where the rotation angles are discretized to mimic
an isotropic distribution of rays (with fewer azimuthal angles close
to the poles)

\[
\theta_{l}=\pi \left({l-1\over N_\theta-1}-{1\over 2}\right),~~
1\le l\le N_\theta,
\]

\begin{equation}
\label{eq:noag}
\phi_{lm}={2\pi m\over N_\phi},~~
1\le m\le N_\phi,~~
N_\phi=2N_\theta\cos\theta_{l}.
\end{equation}

Only those rays which pass through the volume (and not all of them
do!) are stored in memory. Let the total number of all possible
orientations of the base grid be

\begin{equation}
\label{eq:nangles}
N_{\rm angles}\sim{4\over\pi}N_\theta(N_\theta-1).
\end{equation}

\noindent
The resulting number of rays is significantly smaller than $N^3\times
N_{\rm angles}$. In fact, with $N=64$ and $N_{\rm angles}=110$
($N_\theta=10$), we only require $218{,}242$ rays.

Now, that we have the grid of rays and the 3D rectangular mesh, we
have to specify the rules of interpolation between them. Before we
start our simulations, for each 3D grid point $(i,j,k)$ we also store
an array of the closest four rays going through its neighbourhood in
the direction $(\theta_{l},\phi_{lm})$.
Since the rays do not pass exactly through 3D grid points, we use the
values of the intensity on the four closest rays in that direction to
compute the angular-dependent intensity. Assume that for the point
$(i,j,k)$ the distances to the four closest rays going in the
direction $(\theta_{l},\phi_{lm})$ are $d_1$, $d_2$, $d_3$ and $d_4$,
respectively.  We project the point $(i,j,k)$ onto these rays and read
the values of the intensities, which we write as $I_1$, $I_2$, $I_3$
and $I_4$.  We then calculate the intensity at $(i,j,k)$ in the
direction $(\theta_{l},\phi_{lm})$ according to

\begin{equation}
\label{eq:four_intensities}
I_{i,j,k,l,m}=
{\sum_{q=1}^4{d_1 d_2 d_3 d_4\over d_q}w_qI_q}
\bigg/
{\sum_{q=1}^4{d_1 d_2 d_3 d_4\over d_q}w_q},
\end{equation}

\noindent
where the weights $w_q$ for $q>q_0$ are set to zero if $q_0<4$ rays
were found in the immediate neighbourhood of $(i,j,k)$. This might be
the case close to the edges of the computational volume; we shall
comment more on this while discussing the boundary conditions. The
form of eq.~(\ref{eq:four_intensities}) was chosen specifically
because: (1) if a ray labeled $q$ happens to pass exactly through the
point $(i,j,k)$, then $I_{i,j,k,l,m}=I_q$; (2) if all four rays
encompass the point, $\min (I_q)\le I_{i,j,k,l,m}\le \max (I_q)$; and
(3) if $(i,j,k)$ happens to be far from all four rays, then the
resulting intensity will just be an average of the four $I_q$'s.

At each point on our 3D rectangular mesh we assume a piece-wise linear
dependence of the intensity $I$ on two angles, $\theta$ and $\phi$,

\begin{eqnarray}
\label{eq:pwa}
\lefteqn{I_(\theta,\phi)=(1-\xi_\theta)(1-\xi_\phi)I_{i-1,j-1}+{} }\nonumber\\
&&(1-\xi_\theta)\xi_\phi I_{i-1,j}+
\xi_\theta \xi_\phi I_{i,j}+\xi_\theta(1-\xi_\phi)I_{i,j-1},
\end{eqnarray}

\noindent
within a spherical rectangular element (or a spherical triangle
adjacent to either of the poles) bounded by the angles $\theta_l$,
$\theta_{l-1}$ in $\theta$ and $\phi_n$ and $\phi_{n-1}$ in $\phi$,
with the rectangular grid defined as

\[
\theta_{l}=\pi \left({l-1\over N_\theta-1}-{1\over 2}\right),~~
1\le l\le N_\theta,
\]

\begin{equation}
\label{eq:rag}
\phi_n={2\pi n\over N_\theta},~~
1\le n\le N_\theta,
\end{equation}

\noindent
and

\[
\xi_\theta={\theta-\theta_{l-1}\over \theta_l-\theta_{l-1}},~~~~~~
\xi_\phi={\phi-\phi_{n-1}\over \phi_n-\phi_{n-1}}.
\]

We then integrate the intensity over $4\pi$ with appropriate weights
to get the scalar radiation energy density at each point $(l,n)$ of
the rectangular spherical grid

\begin{equation}
\label{eq:energy_density}
E^{\rm diff}_{i,j,k}\equiv {1\over c}\int_{4\pi}I_{i,j,k}(\theta,\phi)d\Omega=
{1\over c}\sum_l\sum_n I^r_{l-1/2\atop n-1/2},
\end{equation}

\noindent
where the quadrature terms for the integration

\[
I^r_{l-1/2\atop n-1/2}\equiv
\int_{\theta_{l-1}\le\theta\le\theta_l\atop\phi_{n-1}\le\phi\le\phi_n}
I_{i,j,k}(\theta,\phi)d\Omega
\]

\noindent
are modified to allow for the non-orthogonal angular grid
(eq.~\ref{eq:noag}).




Since the advection part on the left-hand side of eq.~(\ref{eq:rte})
is strictly linear, the simplest way to propagate intensities is just
to shift wavefronts by one grid zone at each time-step, accounting for
sources and sinks of radiation. Assuming that all discretization
points along each ray are strictly equidistant, the intensity at a
point $j$ is updated simply as

\begin{equation}
\label{eq:wavefront_shift}
I_j^{n+1}=I_{j-1}^n e^{-\kappa_j c\Delta t}+\epsilon_j c\Delta t.
\end{equation}

Alternatively, one could take special care of the length of each ray
segment contained within a particular 3D grid cell, and use a scheme
similar to the third-order-accurate piecewise parabolic advection
method (PPA) of Stone \& Mihalas (1992). In either case we can track
sharp discontinuities in 1D with very little numerical diffusion, and,
therefore, our approach is well suited to the calculation of I-fronts.

\subsection{Local chemistry equations}


Since in our calculation all advection of radiation variables is
performed explicitly, we can solve NLTE rate equations separately at
each point. This makes it relatively easy to implement an implicit
solver for all atomic and molecular processes.

%
%

To demonstrate the capabilities of explicit advection, instead of
solving the proper chemistry equations for multiple species with
primordial chemical composition, we have here adopted a simple toy
model with just photoionization and radiative recombination in a pure
hydrogen medium. The implicit solution of possibly stiff rate
equations described below can be implemented in a similar manner for
more realistic chemistry models.

The time evolution of the degree of ionization $\xe$ is given simply by

\begin{equation}
\label{eq:rate}
{d\xe\over dt}=(1-\xe)g_{\rm HI}E-\xe^2n_{\rm H}\alpha,
\end{equation}

\noindent
with opacity


\begin{equation}
\label{eq:kappa}
\kappa =(1-\xe)n_{\rm H} \hp\nu_1g_{\rm HI}/c.
\end{equation}

\noindent
Here $g_{\rm HI}$ is the photoionization coefficient, $E$ the energy
density of ionizing radiation and $\alpha$ the recombination
coefficient. The correct expression for the total emissivity
$\epsilon$ of the gas can be obtained by considering conservation of
the thermal energy density $E_{\rm th}$ for matter:

\begin{equation}
\label{eq:mec}
\Delta E_{\rm th}\equiv E\left(1-e^{-c\kappa \Delta t}\right)-4\pi\epsilon\Delta t
=\hp\nu_1n_{\rm H}\Delta \xe,
\end{equation}

\noindent
where all $\Delta$ symbols represent the change of variables during one
time step. The full recombination coefficient

\begin{equation}
\label{eq:alpha}
\alpha = \alpha_1+\alpha_{\rm B}
\end{equation}

\noindent
is the sum of recombination coefficients to the ground state
($\alpha_1$) and to all levels above the ground state ($\alpha_{\rm
B}$, the `case B' recombination coefficient), $\nu_1$ is the frequency
just above the Lyman limit, and we assume that recombinations in Lyman
lines occur on a short timescale compared to $(\xe n_{\rm H}\alpha_{\rm
B})^{-1}$. Similarly, the full emissivity (or gas energy loss through
recombinations) is

\begin{equation}
\label{eq:epsilon}
\epsilon = \epsilon_1+\epsilon_{\rm B},
\end{equation}

\noindent
where $\epsilon_1/\epsilon=\alpha_1/\alpha$.
This simple notation ensures radiation energy conservation
in eq.~(\ref{eq:rte}) for pure scattering of Lyman continuum photons
(i.e., when $\alpha_{\rm B}=0$).  Eq.~(\ref{eq:rate}) does not account
properly for the number of photons entering the volume, so that a
large photoionization coefficient $g_{\rm HI}$ might lead to
overproduction of ions. To compensate for this, the number of
photoionizations inside a 3D grid cell
per unit time is not allowed to be larger than the number of photons
actually absorbed inside this cell, i.e.

\begin{equation}
\label{eq:lpi}
{d\xe\over dt}\le {E\over \hp\nu_1}{1-e^{-c\kappa \Delta t}\over n_{\rm H}\Delta t}.
\end{equation}

Eq.~(\ref{eq:rate},\ref{eq:lpi}) are solved separately at each point, given
the local radiation energy density $E$. Discretization of eq.~(\ref{eq:rate})
in time yields

\begin{equation}
\label{eq:fd}
{\xe^{n+1}-\xe^n\over\Delta t}=(1-\theta)f_1(\xe^n)+\theta f_1(\xe^{n+1}),
\end{equation}

\noindent
where $f_1(\xe)$ is just the right-hand side of eq.~(\ref{eq:rate}) and
$1/2<\theta \le 1$ for stability. This equation can almost always be
solved via Newton's method for small enough $\Delta t$.
Linearizing eq.~(\ref{eq:fd}), we get the $(i+1)$-th approximation to
the value of $\xe$ at time $t^{n+1}$:

\begin{eqnarray}
\label{eq:implicit}
\lefteqn{\xe^{n+1,i+1} = \xe^{n+1,i}+
\left[{\partial f_1(\xe^{n+1,i})\over\partial \xe^{n+1,i}}-{1\over\theta\Delta t}\right]^{-1}
\times } \nonumber\\
&&\left[{1-\theta\over\theta}f_1(\xe^n)+f_1(\xe^{n+1,i})-{\xe^{n+1,i}-\xe^n\over\theta\Delta t}\right],
\end{eqnarray}

\noindent
which can then be iterated.

\subsection{The algorithm}

We start calculations by specifying the initial conditions
(temperature, degree of ionization, and in the simplest cases no
radiation field inside the volume) and boundary conditions (the
intensity of radiation entering the volume -- all outward flux at the
edges can freely escape the computational box). The inward flux at the
boundaries is isotropic within $2\pi$ and is simply

\[
I^+={c\over 4\pi}E_{\rm b},
\]

\noindent
where $E_{\rm b}$ is the average background radiation energy density. Since
each ray within the volume starts and ends at the sides of the volume,
we automatically have boundary conditions for each of the 1D advection
problems.
The density field is kept static for all tests in this study, but
since the radiation field is being evolved explicitly at the speed of
light, one could easily evolve the underlying density distribution on
a much bigger fluid flow timescale if desired.

The course of the algorithm at each time step can be divided into
the following steps:

\begin{itemize}
\item project $\epsilon_{\nu}$, $\kappa_{\nu}$ from 3D grid to rays,
\item evolve wavefronts along individual rays by $\Delta t$,
\item project intensities from rays to 3D to reconstruct  $I_{\nu}$
as a function of five variables $I_{\nu}(\bmath{r},\theta,\phi)$,
\item calculate the radiation energy density $E(\bmath{r})$,
\item solve all chemical equations via an implicit iterative technique,
separately at each point, and
\item compute new $\epsilon_{\nu}$, $\kappa_{\nu}$.
\end{itemize}

At the beginning of each time-step we advect 1D intensities
according to eq.~(\ref{eq:wavefront_shift}) along each ray.
The numerical resolution along each ray is simply set to the resolution
$1/N$ of the 3D rectangular mesh,
so that along the ray $i$ the point $j$ has a coordinate

\[
\lambda_j={j-1/2\over N},~~~j=1,...,N_i,~~~N_i=l_iN,
\]

\noindent
where $l_i$ is the length of the ray segment inside the cube.  Note
that one can have much coarser 1D grids along rays, speeding up the
advection but sacrificing both spatial and angular resolution.  The
advected intensities $I_j^{n+1}$ are then projected onto a 3D grid to
reconstruct the mean energy density at each point using
eq.~(\ref{eq:four_intensities} -- \ref{eq:energy_density}). This
operation is one of the most demanding from the computational point of
view, since at each of our $N^3$ points we have to deal with the
angular dependence of the radiation field. When this update is done,
we solve the matter-radiation interaction equations implicitly to
compute the local level populations (eq.~(\ref{eq:implicit})).  This
gives us the 3D distributions of emissivity and opacity
(eq.~(\ref{eq:kappa} -- \ref{eq:epsilon})) which are then mapped back
to the rays and used in the advection scheme at the next time-step.

This simple scheme which we will refer to as {\it time-dependent ray
tracing} can be used as a stand-alone solver, or as a closure scheme
for the system of moment equations through the use of variable
Eddington factors (as in Stone et al.~1992). In the absence of any
sinks and sources of radiation, the intensity is conserved exactly
along each ray. Since the number of rays does not vary with time, our
method guarantees exact conservation of the radiation energy in 3D.

Note that if I-fronts do not propagate fast, for instance, in the
cosmological context where the evolution on the light-crossing
timescale will normally require many thousand time steps, it is
possible to update the radiation energy density $E(\bmath{r})$ once every
few tens or few hundred time steps, while still evolving the
intensities and solving all rate equations properly at the speed of
light.  We found that in practice this shortcut leads to an increase
in speed by a factor of 10 or even higher without any loss of
accuracy. However, it is important to compute the energy density
properly along the edges of I-fronts to guarantee the correct rate of
growth of ionized regions.

To further accelerate this computation, we use adaptive time stepping
to put higher time resolution on the 3D cells in the vicinity of
I-fronts. Since wavefronts cannot propagate further than one grid zone
during one time step, only those cells which just experienced large
change in $E(\bmath{r})$ and their immediate adjacent neighbours need a
proper update of $E(\bmath{r})$. Depending on the width of I-fronts,
typically only a few percent of all 3D cells require the new, exact
value of $E(\bmath{r})$.

One advantage of the use of angle-averaged moments of radiation is
that the advection mechanism is essentially reduced to 3D, and it is
relatively straightforward to implement the multi-dimensional
conservation scheme for the linear advection part of the moment
equations.
Then one could use a much denser spatial grid for the solution of the
moment equations, and a relatively course grid for the angular
reconstruction of the intensity of radiation via ray tracing. In
practice, however, we have found that the mismatch between the spatial
resolutions of the moment solver and of the ray tracing usually leads
to numerical instabilities.  In what follows, we consider ray tracing
only as a stand-alone solver.



Another -- perhaps, a better -- way of coupling angular and spatial
variations of the intensity may be an extension of the Spherical
Harmonics Discrete Ordinate Method \cite{evans98}. For steady-state
transfer problems, instead of storing the radiation field, this method
keeps track of the source function as a spherical harmonic series at each
point. Although the direct implementation of this technique for
time-dependent problems is probably not realistic, due to the lookback
time (i.e. the finite speed of light propagation), the spherical
harmonic representation of the radiation field might require less
storage and might result in smoother angular dependence as compared
with a pure ray tracing approach.

\section{Tests}
\label{sec:tests}

For all of our test runs, except the study of the shadow behind a
neutral clump in Sec.~\ref{sec:shadows}, we set up a numerical grid
with dimensions $64^3\times 10^2$. The angular resolution $N_{\rm
angles}^2\sim 10^2$ was chosen to match the equivalent resolution of
$64^5$ data points for 5D advection. There are $N_{\rm angles}^2$ rays
passing in the immediate neighbourhood ($1/64^3$th of the total
computational volume) of each 3D grid cell, each ray containing
$N_{\rm p}\approx 2/3 \times 64$ grid nodes. Thus, in 5D we obtain the
equivalent resolution of $N_{\rm p}\times 64^3\times N_{\rm
angles}^2\sim 1.1\times 10^9 \sim 64^5$ data points.

Also, for all simulations we take a $2.5\mpc$ (comoving) cosmological
volume, scaled down to $z=10$. All densities (from low-density
optically thin ambient gas to dense clumps) fall in the range
$\Omega_{\rm b}=0.001-0.05$, assuming a Hubble constant of $50 \km
{\rm\,s}^{-1}\mpc^{-1}$. This low density contrast is chosen to cover
the typical range encountered in cosmological hydrodynamics, and is
ideally suited to demonstrate transient features during patchy
ionization. The absorption coefficients are taken to mimic complete
self-shielding (assumed to be $\tau \gsim 10$) at neutral hydrogen
column densities higher than $10^{18}\cm^{-2}$, except where we probe
different regimes, specifically in Sec.~\ref{sec:UV_background} where
$\tau =10$ corresponds to $10^{17}\cm^{-2}$, and
Sec.~\ref{sec:shadows} where $\tau =10$ corresponds to
$10^{21}\cm^{-2}$.

\subsection{An isolated spherical expanding I-front}



A simple problem that would test the ability of the method to track
I-fronts properly is that of a single, isolated Str\"omgren sphere
expanding around a source of ionizing radiation (see, e.g., Abel et
al.~1998).  One difficulty of the current approach is that rays are
drawn in a way to cover the whole computational volume uniformly and
isotropically. For a single point source of radiation this would mean
that only a small number of rays pass through its neighbourhood.
Although intensities along individual rays are strictly conserved,
there is no guarantee that the energy density has the right value far
from a source of a specified luminosity.

At time $t=0$ we turn on a point source, which starts to blow an
expanding $\hii$ bubble around it. Since our algorithm conserves 1D
intensities exactly, independently of spatial or angular resolution,
the I-front must propagate at the right speed, which can be obtained
analytically by equating the number of direct ionizing photons to the
flux of neutral atoms crossing the front. In the absence of radiative
recombinations ($\alpha=0$), the $\hii$ bubble grows indefinitely with
the radius

\begin{displaymath}
\label{eq:pcons}
R_{\rm I}(t)=\left\{\begin{array}{ll}
ct & \textrm{if $t\le t_c$}\\
\left[3\dot{N}_{\rm ph}(t-{2t_c\over 3})/4\pi n_{\rm H}\right]^{1/3} & \textrm{if $t>t_c$},
\end{array} \right.
\end{displaymath}

\noindent
where $t_c=\left({\dot{N}_{\rm ph}/4\pi c^3n_{\rm H}}\right)^{1/2}$.
Parameters used for this calculation are the diffuse neutral gas
density $\Omega_{\rm b}=0.01$ (at $z=10$) and the central source
luminosity $\dot{N}_{\rm ph}=3\times 10^{53} {\rm s}^{-1}$ (all
emitted in photons above the hydrogen Lyman limit). The comparison
between the numerical speed of the I-front and the exact solution from
eq.~(\ref{eq:pcons}) is plotted in Fig.~1. The difference between the
two always stays within one grid zone.


\begin{figure}
\begin{center}
\includegraphics[angle=0,width=7cm]{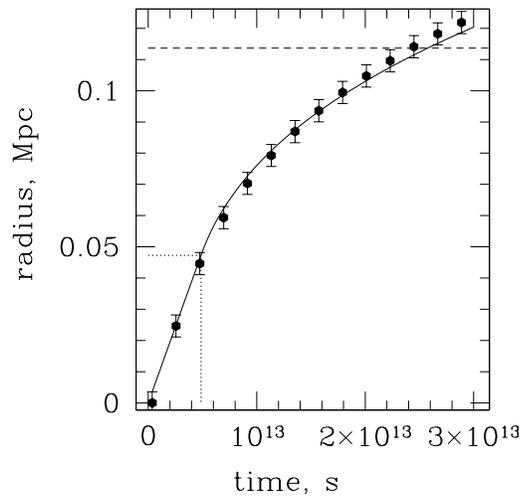}
\caption{The analytical radius of the spherical $\hii$ bubble (solid
line) compared to the numerical results (filled data points). The
error bars indicate the resolution of the grid ($\pm \Delta x$), while
the dotted lines give the radius at which the speed of the I-front
drops below the speed of light. The horizontal dashed line at the top
shows the radius at which the I-front reaches the boundary of the
computational volume.}
\end{center}
\label{fig:front_speed}
\end{figure}

\subsection{An isolated Str\"omgren sphere in the presence of a density
gradient}


For this test we put a point source of radiation into a density
gradient along one of the principal axes of the cube. In the absence
of diffuse radiation from
\hii regions ($\alpha_1=0$) the only ionizing photons come directly
from the source in the centre, in which case the shape of the ionized
bubble would be a simple superposition of Str\"omgren spheres with
radii $R_{\rm S}(\phi)$ varying with the azimuthal angle $\phi$ and
given by the classical solution \cite{spitzer68}

\begin{equation}
\label{eq:stromgren_radius}
S_0=4\pi \alpha_{\rm B}\int_0^{R_{\rm S}(\phi)}r^2n_{\rm H}^2(r,\phi)dr,
\end{equation}

\noindent
where $S_0$ is the photon production rate of the central source.
For an exponential density gradient along the y-axis

\begin{equation}
\label{eq:density_gradient}
\log n_{\rm H}=\log n_{\rm H,1}+{r\cos \phi +y_0\over \Delta y}
\log \left(n_{\rm H,2}\over n_{\rm H,1}\right)
\end{equation}

\noindent
($n_{\rm H,1}$, $n_{\rm H,2}$ being the hydrogen densities on the opposite
faces of the cube), the equilibrium Str\"omgren radius $R_{\rm
S}(\phi)$ is given by a simple equation

\begin{equation}
\label{eq:stromgren_radius_exponential_gradient}
S_0={4\pi \alpha_{\rm B}\over b^3}\left[e^{a+bR_{\rm S}(\phi)}(b^2R_{\rm S}^2(\phi)-2bR_{\rm S}(\phi)+2)-2e^a\right],
\end{equation}

\noindent
where


\[
a\equiv 2\left[\ln n_{\rm H,1}+{y_0\over \Delta y}\ln \left(n_{\rm H,2}\over n_{\rm H,1}\right)\right]
~~{\rm and}~~~
\]
\[
b\equiv {2\cos \phi \over \Delta y}\ln \left(n_{\rm H,2}\over n_{\rm H,1}\right).
\]

We take the physical densities on the opposite faces of the volume to
be $\Omega_{\rm b,1}=0.001$ and $\Omega_{\rm b,2}=0.05$, and the
luminosity is $\dot{N}_{\rm ph}=10^{51} {\rm s}^{-1}$. Since we do not
solve any realistic atomic rate equations in this paper, we take the
value of the total hydrogen recombination rate from Hummer (1994) for
some fiducial temperature ($T=10^4 {\rm K}$). In Fig.~2 we plot a time
sequence of models, for ionization by a central source, with no
scattering of Lyman photons (i.e. $\alpha_1=0$).
The numerical solution at $t=4.1~{\rm Gyrs}$ appears to be very close
to the exact one for an equilibrium Str\"omgren sphere. The sharp
transition layer between the ionized and the neutral regions in the
high optical depth regime indicates that, indeed, the scheme
introduces very little numerical diffusion even when extended to 3D.


\begin{figure*}
\begin{center}
\includegraphics[angle=-90,width=14.5cm]{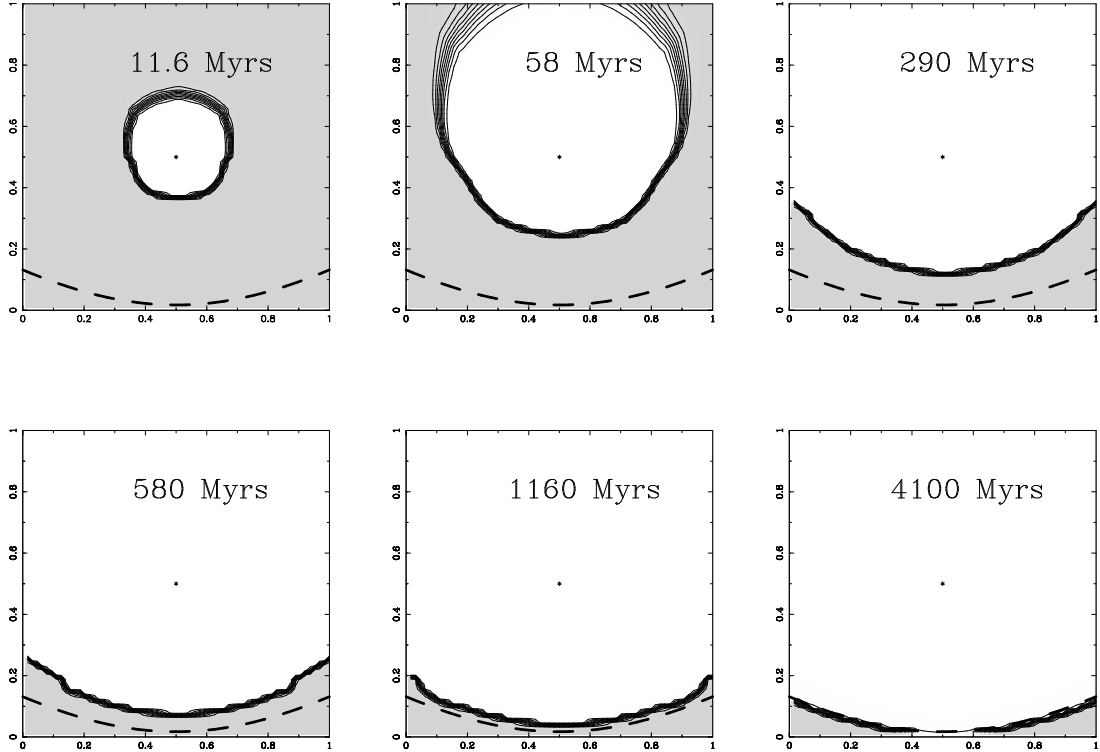}
\caption{ The cross-section of the numerical 3D ionization front going
through the source of ionization for a model with no scattering, shown
for six output times. The density of the gas follows an exponential
gradient (eq.~\ref{eq:density_gradient}) increasing by a factor of
$50$ from the upper to the lower side of the cube. The dashed line
gives the analytic solution
(eq.~\ref{eq:stromgren_radius_exponential_gradient}) for an equilibrium
Str\"omgren sphere without scattering. The width of the transition
layer between the neutral and fully ionized regions is consistent
everywhere with the mean free path of ionizing photons. The ionizing
source of luminosity $\dot{N}_{\rm ph}=10^{51} {\rm s}^{-1}$ (all
emitted in photons just above the hydrogen Lyman limit) is marked by
the asterisk.
The nine contour levels correspond to ionization fractions of
$\xe=0.1,0.2,...,0.9$.}
\end{center}
\label{stromgren_gradfig}
\end{figure*}

\begin{figure}
\begin{center}
\includegraphics[angle=0,width=7cm]{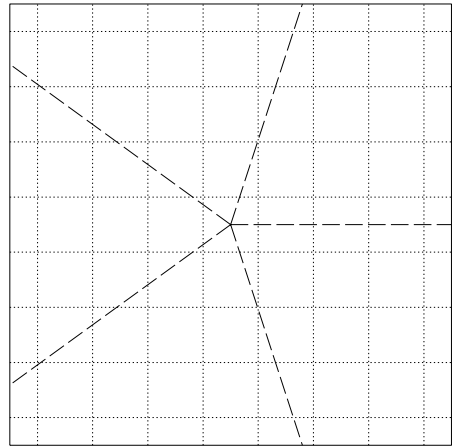}
\caption{In order to cover the whole computational volume with rays
uniformly and isotropically, we set the number of rays inside a
latitudinal angle interval $[\theta,\theta+d\theta]$ proportional to
$\cos \theta$. An odd number of rays will necessarily introduce slight
left-right asymmetry for interpolation between the 3D rectangular grid
and the radial grid for any non-central cross-section through the
volume. In this figure we show a 2D schematic representation of the
two grids employed in our method. To simplify the plot, the source of
radiation is assumed to be sitting in the centre, and only radial rays
are drawn here. Due to the finite angular resolution, interpolation
between the 3D mesh and the rays is not symmetric with respect to the
centre. This shows up in the asymmetry between the upper and the lower
sides of the ionization contours in Fig.~8.}
\end{center}
\label{two_grids}
\end{figure}


\subsection{Ionization in the presence of a UV background}
\label{sec:UV_background}


The uniform coverage of the whole volume with rays implies that
extended sources of radiation will be represented statistically much
better than point sources. A simple test mimicking the evolution of
dense clouds in the presence of ionizing radiation is to enclose the
computational region in an isotropic bath of photons. The simplest way
to accomplish this is just to set up a uniform, isotropically glowing
boundary at the edges of the cube at $t=0$. An effective demonstration
of time-dependent ray tracing would be its ability to deal with any
distribution of state variables within the simulation volume. For this
test, we set up a density condensation shaped as the acronym for
`radiation hydrodynamics' (RHD), with a density $50$ times that of the
ambient homogeneous medium. The ambient medium has a constant density
of $\Omega_{\rm b}=10^{-3}$, and the energy density of the background
radiation is $E=5\times 10^{-23}\nu_1\erg\cm^{-2}\hz^{-1}\ster^{-1}$.
Fig.~4 shows the result of this run. Most of the low-density
environment is ionized on the radiation propagation timescale. It
takes somewhat longer for ionizing photons to penetrate into the dense
regions. Whether these regions can be ionized on a timescale of
interest, depends on the ratio of the recombination timescale to the
flux of background radiation.
One can easily see ionization `eating in' to the neutral zone, e.g. in
the disappearance of the serifs on the letters at late times. Note
that the width of the ionization fronts does not usually exceed one
grid zone (Fig.~5).

\begin{figure*}
\begin{center}
\includegraphics[angle=0,width=14.5cm]{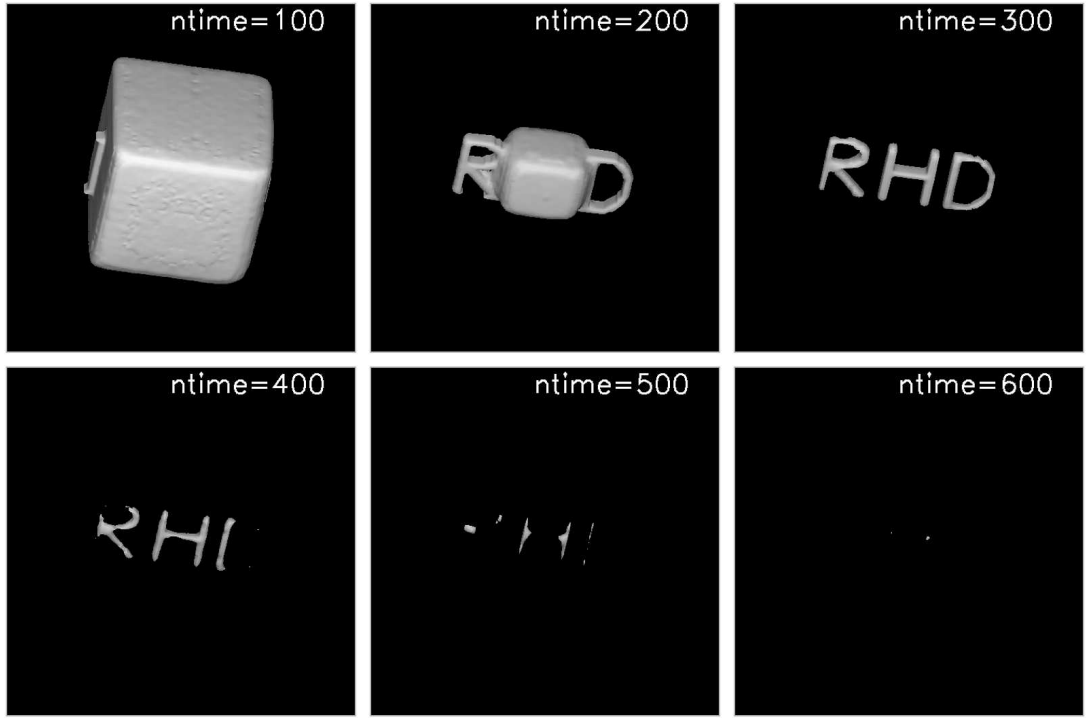}
\caption{
The isosurface of ionization level $\xe=0.5$ is plotted at six
time intervals for the model with an ionizing background coming from
outside of the cube. The density contrast between the ambient medium
and the high-density acronym `RHD' (radiation hydrodynamics) is
$50$. This simulation demonstrates 5D advection on the radiation
propagation time-scale. The numerical resolution is $64^5$.
The effects of shielding are clearly visible during partial
ionization.}
\end{center}
\label{rhdfig}
\end{figure*}

\begin{figure*}
\begin{center}
\includegraphics[angle=-90,width=14.5cm]{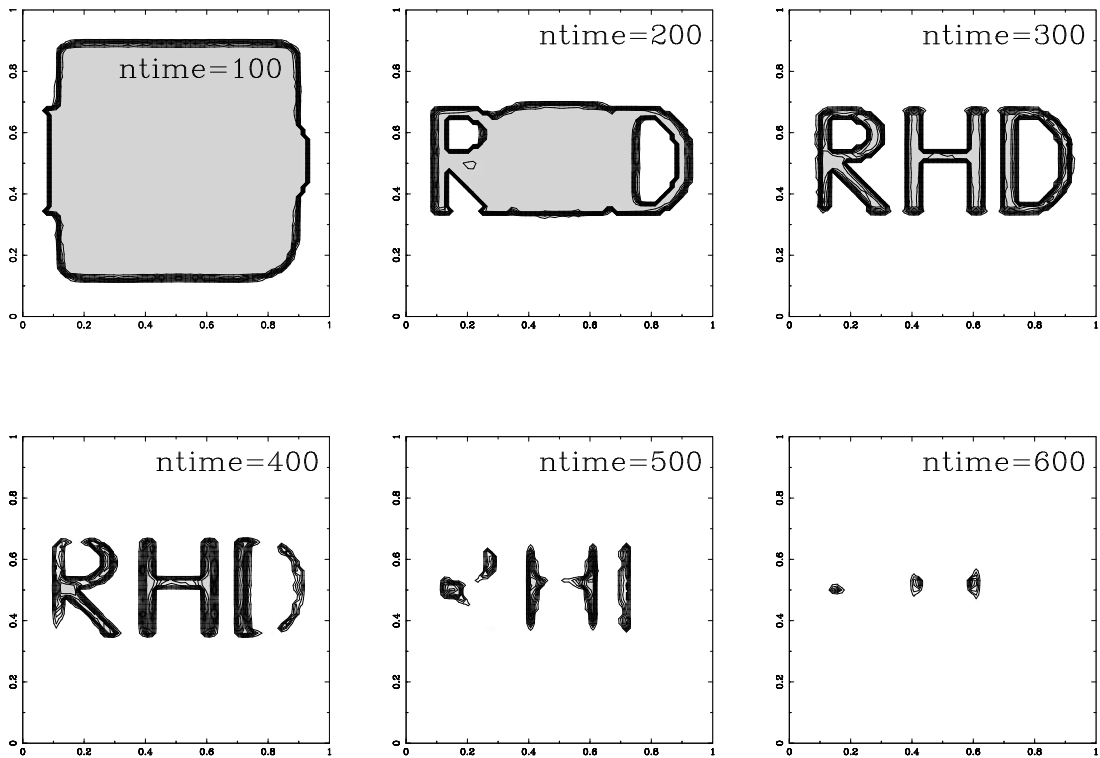}
\caption{ Contour plot of ionization in a cross-section through the
data presented in Fig.~4. There are 11 contour lines spaced linearly
from $\xe=0$ to $\xe=1$. Note the sharpness of the ionization
boundaries, the shielding, and the fact that the most deeply embedded
regions take the longest to ionize.}
\end{center}
\label{rhd_slicefig}
\end{figure*}

\subsection{Diffuse radiation from \hii regions: shadows behind neutral clouds}
\label{sec:shadows}


Part of the ionizing radiation at high redshifts comes in the form of
hydrogen Lyman continuum photons from recombinations in diffuse
ionized regions. The following test, simulating the formation of
shadow regions behind dense clouds at the resolution $32^3\times
10^2$, was adapted from Canto, Steffen \& Shapiro (1998). A neutral
clump of radius $R_{\rm c}=1.25\mpc$ (comoving) is being illuminated
by a parallel flux $F_*=6.34\times 10^{-16}\hp^{-1}\erg
\cm^{-2}\s^{-1}\hz^{-1}$ of stellar ionizing photons (just above
$13.6\ev$) from one side; here $\hp$ is Planck's constant.  A shadow
behind the clump is being photoionized by secondary recombination
photons from the surrounding \hii region (Fig.~6) of physical density
$n_{\rm H}=3.7\times 10^{-2}\cm^{-3}$. Neglecting hydrodynamical
effects, the width $R_{\rm I}$ of the shadow region can be estimated
using a simple two-stream approximation \cite{canto98}:

\begin{equation}
\label{eq:two_stream_approximation}
\xi^2={2\over 1+r_i^2(2\ln r_i-1)}, ~~~{\rm where} ~~~r_i\equiv {R_{\rm I}\over R_{\rm c}},
\end{equation}

\noindent
and the dimensionless parameter $\xi$ is defined as

\begin{equation}
\label{eq:xi_parameter}
\xi \equiv {4R_{\rm c} n_{\rm H}^2\over F_*}{\alpha_{\rm B}^2\over \alpha_1}.
\end{equation}

For $\xi^2\le 2$, recombination Lyman continuum photons from the
illuminated region will eventually photoionize the shadow completely.
For $\xi^2>2$, radiative losses through low-energy cascade
recombination photons will stop the I-front, forming a neutral
cylinder behind the dense clump. Strictly speaking, equations
(\ref{eq:two_stream_approximation})--(\ref{eq:xi_parameter}) are valid
only for a shadow completely photoionized by secondary photons, and
should be viewed as an approximation to I-fronts driven by scattering.

For this run we modify boundary conditions to include recombination
photons originating outside the box. Each of the rays -- starting on
any face except the upper side of the volume (which goes through the
neutral shadow) -- carries the additional intensity of
$\left\langle\epsilon_1\right\rangle/\left\langle\kappa\right\rangle$,
where the angular brackets denote the average throughout the currently
photoionized gas inside the box.


In Fig.~7 we plot the radius $R_{\rm I}$ of the shadow neutral region
as a function of $\xi$ in our 3D numerical models. To play with the
size of the neutral core, we fix the total recombination coefficient
but we vary the portion of recombinations into the ground state (which
would produce more Lyman continuum photons capable of ionizing the
medium). The width of the I-front driven by secondary photons depends
on the assumed opacity (the optical depth of $\tau=10$ corresponding
to the column density of $10^{21}\cm^{-2}$). This low opacity was
chosen to reach equilibrium quicker -- equilibrium itself does not
depend on the opacity -- but it cannot be too low otherwise there will
be an unnecessary spread of the transition zone in the I-front over
many grid cells.

We find a remarkably good agreement between the results of our models
and the analytic solution, taking into account that ionization of the
shadow is due to scattering in the medium.


\begin{figure*}
\begin{center}
\includegraphics[angle=-90,width=14.5cm]{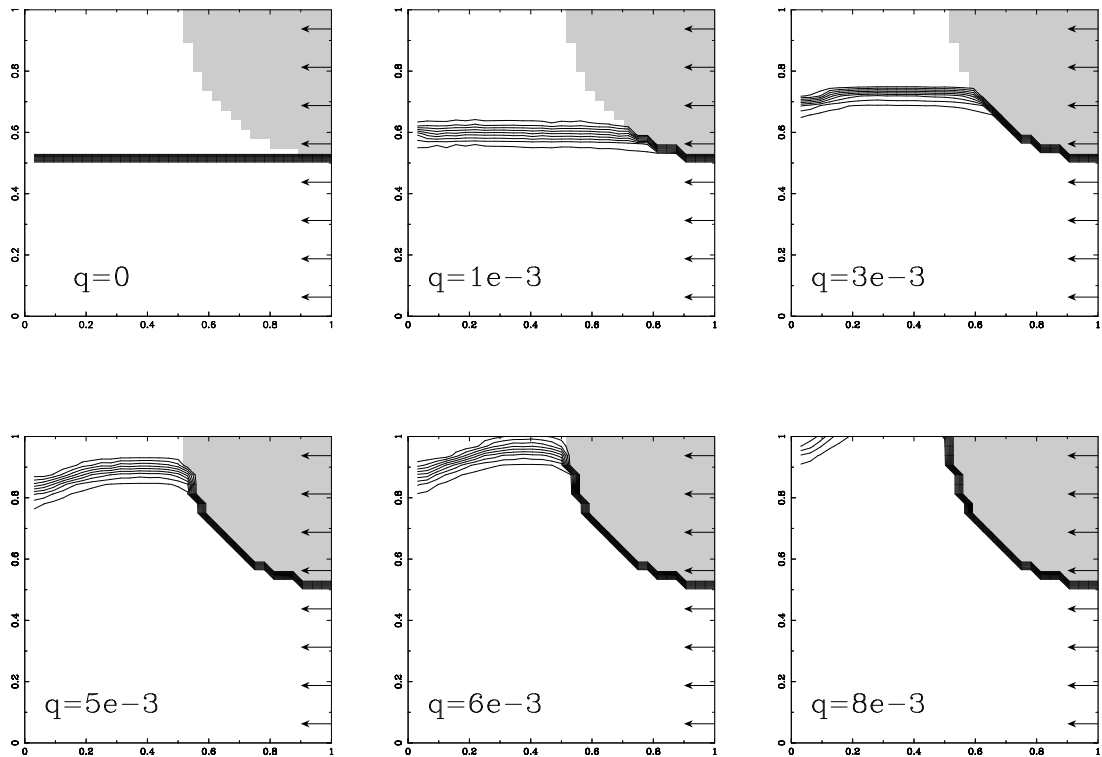}
\caption{ Cross-sections of the dense clump (shaded) and the neutral
shadow region behind it, for different fractions $q=\alpha_1/\alpha$
of recombination into ground states. The clump is being illuminated
with a parallel flux $F_*$ of direct stellar ionizing photons entering
through one side of the box (marked with arrows), and the shadow
region behind it is being photoionized by recombination Lyman
continuum photons from the surrounding \hii gas. Each model has been
evolved to an equilibrium state, correspoding to many passages of the
wavefront across the computational region. The nine contour levels
correspond to ionization fractions of $\xe=0.1,0.2,...,0.9$.
This test is adapted from Canto et al.~(1998).}
\end{center}
\label{shadow_fig_1}
\end{figure*}

\begin{figure}
\begin{center}
\includegraphics[angle=0,width=7cm]{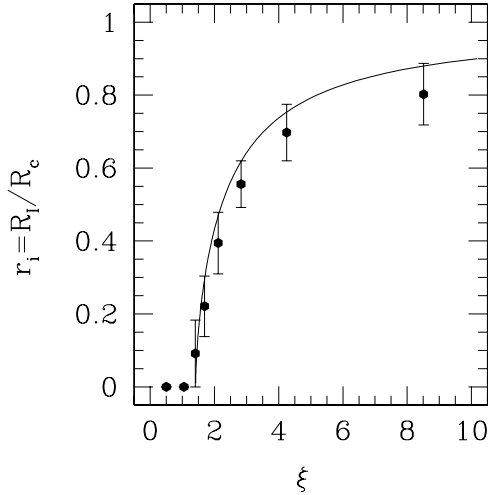}
\caption{The width $R_{\rm I}$ of the shadow behind a neutral clump of
radius $R_{\rm c}$ as a function of the dimensionless parameter $\xi$
of eq.~(\ref{eq:xi_parameter}). The solid line represents the
two-stream analytic solution (Canto et al.~1998), assuming an
infinitely sharp boundary. The points show the results of our 3D
numerical model, plotted explicitly at $\xe=0.5$, with the errorbars
giving the width of the I-front formed by secondary photons. Since the
width of the neutral core in our numerical model depends on multiple
scatterings in 3D,
we would not expect a detailed match; however, the agreement between
the approximate analytic and the numerical solutions is remarkably
good.}
\end{center}
\label{shadow_fig_2}
\end{figure}

\subsection{Diffuse radiation from \hii regions:
ionization of a central void}


To demonstrate the ability of our scheme to handle scattering in more
complicated situations, we also set up a model with ionization of a central
low-density void by secondary, recombination photons. The void region is
surrounded by two nested cubes with opposite faces open. The walls of the
cubes are set to be much denser than the rest of the medium, to screen
completely the central void from direct ionizing photons, and the ionizing
UV flux is introduced at all faces of the computational volume. The total
hydrogen recombination coefficient $\alpha_1+\alpha_{\rm B}$ is again taken
for the temperature $T=10^4{\rm K}$. Similar to Sec.~4.4, we vary the amount
of scattering in the medium by changing the fraction $q$ of atoms
recombining into the ground state. In reality at $T=10^4{\rm K}$ the value
$q\approx 0.38$ \cite{hummer94} gives a solution in between our extreme
values of $q$.

\begin{figure*}
\begin{center}
\includegraphics[angle=-90,width=14.5cm]{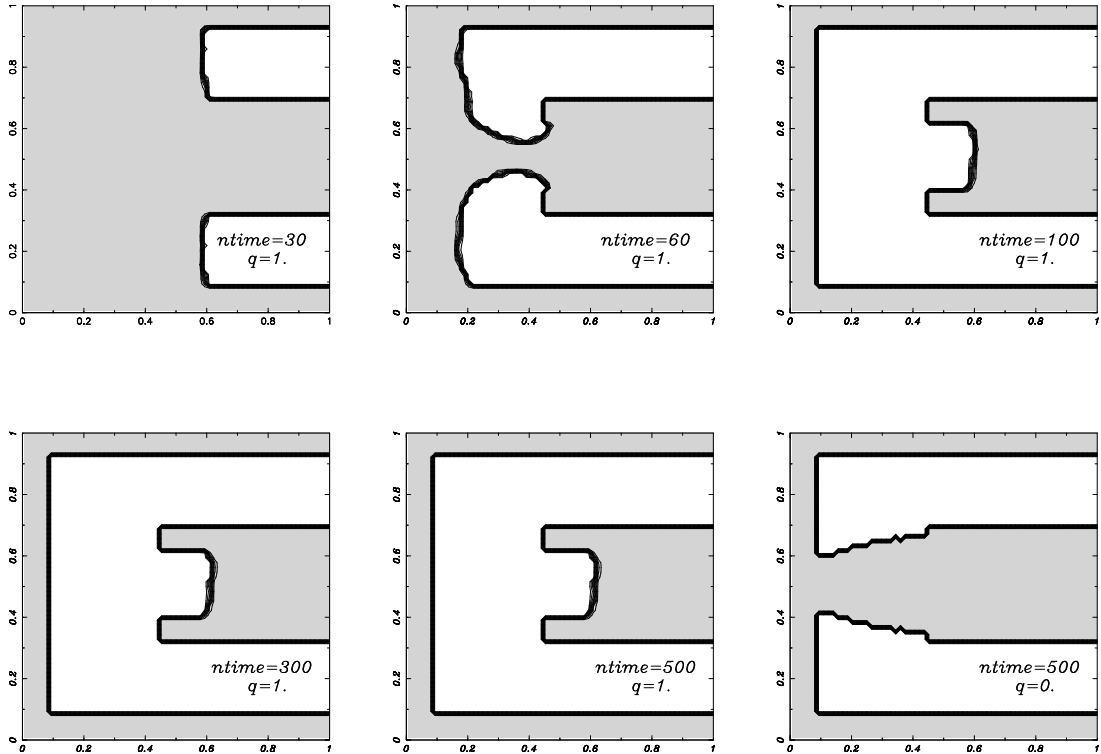}
\caption{ Diagram demonstrating ionization of the void region, which
consists of two nested cubes with opposite sides open. The central
void and the side `tunnels' have the same density as the surrounding
medium, and the density of the walls is set to be much higher to
completely block all external ionizing photons. In this figure we
plot one of the central cross-sections showing contours of ionization
at five different output times for the model with complete scattering
($q=1$), and a quasi-equilibrium configuration for the model with no
scattering ($q=0$) after $n=500$ timesteps. The shaded arears represent
neutral regions.}
\end{center}
\label{void_fig}
\end{figure*}

Similar to the test problem of Section~\ref{sec:UV_background}, if
$\alpha_{\rm B}=0$, then the medium will be ionized completely, since
there is a constant flux of primordial ionizing photons. The speed of
ionization depends on the values of $\alpha_1$, $\alpha_{\rm B}$,
$g_{\rm HI}$ and $n_{\rm H}$. Note, however, that if $\alpha_1$ is too
high, the I-front will be very slow, since a large portion of the
original ionizing photons are scattered back. On the other hand, if
$\alpha_1$ is too low, the I-front will propagate much faster in those
regions where ionization is driven by primordial photons, but in
shadowed regions there will be too few recombination photons. Thus, it
seems that the speed of ionization of the central void will be the
highest at some intermediate $\alpha_1$.

In Fig.~8 we demonstrate ionization of the void region for models with
complete scattering and with no scattering at all. The parameters used
for this model are $E=5\times
10^{-20}\nu_1\erg\cm^{-2}\hz^{-1}\ster^{-1}$ and the diffuse neutral
gas density $\Omega_{\rm b}=0.01$.

As expected, for the no-scattering model ($q=0$) the central region
remains neutral, since there is no direct path for the ionizing
photons.  However, for the model which includes scattering ($q=1$), at
least part of the central region becomes ionized. This demonstrates
that our scheme is perfectly capable of dealing with re-scattering of
the ionizing photons. Beyond this simple test case, there are many
astrophysical situations where progress can be made via numerical
radiative transfer. For example, analytic solutions are often used,
which are steady-state, and which assume a sharp boundary between the
neutral and ionized zones.  Using our numerical techniques it should
be possible to follow general systems with complex density
inhomogeneities as well as regions of partial ionization.


\section{Conclusions}
\label{sec:conclude}

In this paper we have demonstrated that, with existing desktop
hardware, it is possible to model cosmological inhomogeneous
reionization on a light-crossing time $t_{\rm R}$ in three spatial
dimensions. Since {\it the photoionization time-scale in the low
optical depth regime ($\tau\leq 1$) is of order of the light-crossing
time $t_{\rm R}$, explicit advection might be a faster method in
covering at least these regions}. Compared to the elliptic-type
solvers on the fluid-flow time-scale or the time-scale of atomic
processes, explicit radiative advection produces very accurate results
without the need to solve a large system of coupled non-linear
elliptic equations. The computing requirements with explicit advection
grow linearly with the inclusion of new atomic and molecular rate
equations, which is certainly not the case for quasi-static solvers
(although it is feasible that the development of multigrid techniques
for elliptic equations might actually approach similar scaling).

Using eq.~(\ref{eq:cfl}) we can see that the entire history of
reionization can be modeled with $\sim10^4$--$10^5$ time-steps
(depending on the required resolution), which makes explicit advection
an attractive choice for these calculations. However, the efficiency
of the explicit radiative solver has still to be explored. Future work
should include a detailed comparison between explicit advection and
implicit reconstruction (through an elliptic solver), to demonstrate
which method works best for calculating inhomogeneous reionization.

As we have demonstrated here, for certain problems, including the
propagation of supersonic I-fronts, the Courant condition does not
seem to impose prohibitively small time-steps. In this case the
biggest challenge is to accurately describe anisotropies in the
radiation field, i.e. to solve for inhomogeneous advection in the 5D
phase space, in the presence of non-uniform sources and sinks of
radiation. Strictly speaking, the storage of one variable at, say,
$64^5$ data points requires about $9\,$GB of memory, which stretches
the capabilities of top-end desktop workstations. One attractive
possibility for future exploration is to directly solve the
monochromatic photon Boltzmann equation in 5D. To demonstrate the
feasibility of the numerical solution, however, among different
methods, we here chose to concentrate on simple ray tracing at the
speed of light. The numerical approach we have used is completely
conservative and produces very little numerical dissipation.

We want to conclude that despite the high dimensionality of the
problem, with a reasonable expenditure of computational resources, of
the type available today, it is possible to numerically model many
different aspects of the full 3D radiative transfer
problem. Furthermore we feel that the methods described here represent
a significant and realizable step towards the goal of full
cosmological RHD.

\section*{ACKNOWLEDGMENTS}
We wish to thank Taishi Nakamoto for providing us with the results of
reionization models from the University of Tsukuba ahead of
publication. A.R. would like to thank Jason R. Auman for numerous
enlightening discussions, and Gregory G. Fahlman for constant
encouragement on this project, as well as Randall J. LeVeque for help
with numerical methods for multidimensional conservation laws. This
work was supported by the Natural Sciences and Engineering Research
Council of Canada.


\label{lastpage}


\begin{thebibliography}{99}

\bibitem[\protect\citename{Abel et al.} 1997]{abel97} Abel T., Anninos P.,
	Zhang Y., Norman M.L., 1997, New Astronomy, 2, 181


\bibitem[\protect\citename{Abel et al.} 1998]{abel98} Abel T., Norman M.L.,
	Madau P., 1998, astro-ph/9812151

\bibitem[\protect\citename{Anninos et al.} 1997]{anninos97} Anninos P.,
	Zhang Y., Abel T., Norman M.L., 1997, New Astronomy, 2, 209

\bibitem[\protect\citename{Auer \& Mihalas} 1970]{auer70} Auer L.H.,
	Mihalas D., 1970, MNRAS, 149, 65

\bibitem[\protect\citename{Canto et al.} 1998]{canto98}
	Canto J., Steffen W., Shapiro P.R., 1998, ApJ, 502, 695

\bibitem[\protect\citename{Couchman \& Rees} 1986]{couchman86}
	Couchman H.M.P., Rees M.J., 1986, MNRAS, 221, 53

\bibitem[\protect\citename{Evans} 1998]{evans98} Evans F., 1998,
	J. Atmospheric Sciences, 55, 429

\bibitem[\protect\citename{Gnedin \& Ostriker} 1997]{gnedin97} Gnedin N.Y.,
	Ostriker J.P., 1997, ApJ, 486, 581


\bibitem[\protect\citename{Haiman \& Loeb} 1997]{haiman97} Haiman Z.,
	Loeb A., 1997, ApJ, 483, 21

\bibitem[\protect\citename{Haiman \& Loeb} 1998]{haiman98} Haiman Z.,
	Loeb A., preprint astro-ph/9710208

\bibitem[\protect\citename{Hummer} 1994]{hummer94} Hummer D.G., 1994,
	MNRAS, 268, 109

\bibitem[\protect\citename{Katz et al.} 1996]{katz96} Katz N.,
	Weinberg D.H., Hernquist L., Miralda-Escud\'e J., 1996, ApJ, 457, L57

\bibitem[\protect\citename{Kepner et al.} 1997]{kepner97} Kepner J.V.,
	Babul A., Spergel D.N., 1997, ApJ, 487, 61

\bibitem[\protect\citename{Madau et al.} 1997]{madau97} Madau P.,
	Meiksin A., Rees M.J., 1997, ApJ, 475, 429

\bibitem[\protect\citename{Madau} 1998]{madau98} Madau P., 1998, preprint
	astro-ph/9807200

\bibitem[\protect\citename{Meiksin} 1994]{meiksin94} Meiksin A., 1994,
	ApJ, 431, 109

\bibitem[\protect\citename{Mihalas \& Mihalas} 1984]{mihalas84} Mihalas D.,
	Mihalas B., 1984, Foundations of Radiation Hydrodynamics.
	Oxford University Press, New York

\bibitem[\protect\citename{Miralda-Escud\'e} 1998]{miralda98} Miralda-Escud\'e J.,
	Haehnelt M., Rees M.J., 1998, astro-ph/9812306

\bibitem[\protect\citename{Navarro \& Steinmetz} 1997]{navarro97} Navarro J.F.,
	Steinmetz M., 1997, ApJ, 478, 13

\bibitem[\protect\citename{Norman et al.} 1998]{norman98} Norman M.L., Paschos P.,
	Abel T., 1998, astro-ph/9807282

\bibitem[\protect\citename{Park \& Hong} 1998]{park98} Park Y.-S.,
	Hong S.S., 1998, ApJ, 494, 605

\bibitem[\protect\citename{Shapiro} 1986]{shapiro86} Shapiro P.R., 1986,
	PASP, 98, 1014

\bibitem[\protect\citename{Shapiro et al.} 1998]{shapiro98} Shapiro P.R.,
	Raga A.C., Mellema G., 1998, preprint astro-ph/9804117

\bibitem[\protect\citename{Spitzer} 1968]{spitzer68} Spitzer L., Jr., 1968,
	Diffuse matter in space. Interscience publishers, New York

\bibitem[\protect\citename{Stone \& Mihalas} 1992]{stone92a} Stone J.M.,
	Mihalas D., 1992, JComputPhys, 100, 402

\bibitem[\protect\citename{Stone et al.} 1992]{stone92b} Stone J.M.,
	Mihalas D., Norman M. L., 1992, ApJS 80, 819


\bibitem[\protect\citename{Tajiri \& Umemura} 1998]{tajiri98} Tajiri Y.,
	Umemura M., 1998, preprint astro-ph/9806046

\bibitem[\protect\citename{Umemura et al.} 1998]{umemura98} Umemura M.,
	Nakamoto T., Susa H., 1998, in Miyama S. M., Shibata K., eds,
	Numerical Astrophysics 1998. Kluwer, in press

\bibitem[\protect\citename{Zhang et al.} 1998]{zhang98} Zhang, Y., Meiksin, A.,
	Anninos, P., Norman, M.L., 1998, ApJ, 495, 63

\end{thebibliography}
\end{document}